\documentclass[aps,prd,nofootinbib,twocolumn,superscriptaddress,preprintnumbers,letterpaper,natbib,nofootinbib]{revtex4-1}
\pdfoutput = 1
\usepackage{amssymb}
\usepackage{amsmath}
\usepackage[usenames,dvipsnames]{color}
\usepackage{hyperref}
\usepackage{graphicx}

\usepackage[page,toc,titletoc,title]{appendix}

\hypersetup{
    colorlinks=true,       
    linkcolor=Cyan,          
    citecolor=Magenta}   
    
\makeatletter

\newcommand{\bea}{\begin{eqnarray}}
\newcommand{\eea}{\end{eqnarray}}

\begin{document}

\title{The first limit on invisible decays of $B_s$ mesons comes from LEP}

\author{Gonzalo Alonso-\'Alvarez}
\email{galonso@physics.mcgill.ca}
\thanks{ORCID: \href{https://orcid.org/0000-0002-5206-1177}{0000-0002-5206-1177}}
\affiliation{McGill University Department of Physics, 3600 Rue University, Montr\'eal, QC, H3A 2T8, Canada}
\affiliation{Department of Physics, University of Toronto, Toronto, ON M5S 1A7, Canada}

\author{Miguel Escudero Abenza}
\email{miguel.escudero@cern.ch}
\thanks{ORCID: \href{https://orcid.org/0000-0002-4487-8742}{0000-0002-4487-8742}}
\affiliation{Theoretical Physics Department, CERN, 1211 Geneva 23, Switzerland}

\begin{abstract} 
\noindent Motivated by the recent evidence for $B^+\to K^+\bar{\nu} \nu$ decays at Belle II, we point out that fully invisible $B_d$ and $B_s$ meson decays are strongly constrained by LEP. A reinterpretation of an old inclusive ALEPH search for $b$-hadron decays with large missing energy allows us to place the limits $\mathrm{Br}(B_d \rightarrow \mathrm{invisible}) < 1.4\times10^{-4}$ and $\mathrm{Br}(B_s \rightarrow \mathrm{invisible}) < 5.6\times10^{-4}$, both at $90\%$ CL. The $B_d$ limit is only a factor of 6 looser than the world-leading one provided by the BaBar collaboration, while the $B_s$ one is the first limit in the literature on this decay mode. These results are relevant in the context of new light states coupled to quarks and exemplify the power of a future Tera-$Z$ factory at FCC-ee to look for $B$ meson decays containing missing energy.
\end{abstract}

\preprint{CERN-TH-2023-193}

\maketitle

{
  \hypersetup{linkcolor=black}
}

\setlength\parskip{4pt}

\section{Introduction}
\vspace{-0.4cm}

Rare decays of $b$-flavored mesons are a promising avenue to look for new physics.
The $b\rightarrow s$ flavor-changing neutral currents, which are strongly suppressed in the Standard Model (SM), have gathered a lot of attention in the past few years.
The transition involving neutrinos, $b\rightarrow s \bar{\nu} \nu$, is theoretically well-suited to test the consistency of the Standard Model~\cite{Buchalla:1993bv,Grossman:1995gt,Buchalla:1995vs,Bartsch:2009qp,Buras:2014fpa} and offers a window to weakly-coupled light new physics~\cite{Bird:2004ts,Altmannshofer:2009ma,Filimonova:2019tuy,MartinCamalich:2020dfe,Browder:2021hbl,Bause:2021cna,He:2021yoz,Felkl:2021uxi,Ferber:2022rsf,He:2022ljo,Ovchynnikov:2023von,Asadi:2023ucx}.
However, the associated $B$ meson decay modes are challenging to test experimentally due to the presence of missing energy in the final state, and until recently only upper limits on the $B\rightarrow K \bar{\nu} \nu$ channels existed~\cite{BaBar:2013npw,Belle:2013tnz,Belle:2017oht,Belle-II:2021rof}.

The experimental situation has very recently been upended by the $3.5\sigma$ detection of ${\rm Br}(B^+\rightarrow K^+ \bar{\nu}\nu) = (2.3\pm 0.7)\times 10^{-5}$ by the Belle II collaboration~\cite{Belle-II:2023esi}.
This rate is $2.7\sigma$ larger than the Standard Model expectation~\cite{Becirevic:2023aov}, which has prompted ample discussion~\cite{Bause:2023mfe,Athron:2023hmz,Allwicher:2023xba,Felkl:2023ayn,He:2023bnk,Dreiner:2023cms,Altmannshofer:2023hkn,McKeen:2023uzo,Bolton:2024egx}.
Of particular interest is the interplay with the other channels triggered by the $b\rightarrow s \bar{\nu} \nu$ transition, $B_d \rightarrow K^{(\star)0} \bar{\nu} \nu$ and $B_s \rightarrow \bar{\nu} \nu$.
Current upper limits on the former mode~\cite{Belle:2017oht} are only a factor of $\sim 2$ away from the SM prediction~\cite{Becirevic:2023aov}, but up to now no bounds exist on the invisible $B_s$ decay rate.
This is a critical channel since an enhanced rate above the extremely suppressed SM prediction~\cite{Lu:1996et,Badin:2010uh,Bortolato:2020bgy} would constitute a smoking gun signal of the presence of light new physics in $b\rightarrow s$ transitions~\cite{Bause:2023mfe,Felkl:2023ayn}. It is therefore very timely to explore the extent to which existing and upcoming experiments can look for fully invisible decays of neutral $B$ mesons. 
In this study, we argue that $e^+e^-$ colliders running at the $Z$ pole are excellently positioned to carry out this task.

The ALEPH collaboration at LEP performed a set of inclusive searches for $b$-quark decays involving large missing energy. 
In particular, the collaboration was able to make the first measurement of $b\to \tau^-\bar{\nu}_\tau X $~\cite{ALEPH:1992zwu}, and set world record limits (at the time) on $B^-\to \tau^- \bar{\nu}_\tau$~\cite{ALEPH:1994bih}. Around the time when these analyses were performed, Grossman, Ligeti \& Nardi pointed out that these type of searches could also be used to constrain inclusive $b\to s\bar{\nu}\nu$ decays~\cite{Grossman:1995gt}, as well as $B\rightarrow X \tau^+\tau^-$ and the radiative $B\rightarrow \gamma \bar{\nu}\nu$ mode~\cite{Grossman:1996qj}, see also~\cite{Mangano:1997md} for $B_c \to \tau \nu$ decays. In fact, a subsequent dedicated analysis by ALEPH~\cite{ALEPH:2000vvi} lead to the still-standing most stringent upper limit on the first of these decay channels: ${\rm Br}(b\to s \bar{\nu}\nu) < 6.4\times 10^{-4}$ at 90\% CL, see also~\cite{ParticleDataGroup:2022pth}.
The strategy behind these searches at ALEPH is simple~\cite{ALEPH:1992zwu,ALEPH:1994bih,ALEPH:2000vvi}: to look for hadronic $Z$ decays with $b\bar{b}$ quarks in the final state where one of these $b$'s produces large amounts of missing energy in its decay. 
The basic prerequisite for the selected events is the presence of two jets in opposite hemispheres.
Once a jet is tagged as arising from a $b$ quark, the missing energy is reconstructed in the opposite one by subtracting the energy deposited in the various calorimeters to the known initial beam energy. 
This renders the search completely inclusive: any event with a $b$ quark whose decay contains sufficient missing energy falls in the signal region. 
Since no possible decay mode leads to more missing energy than a fully invisible $B_s$ or $B_d$ decay, the ALEPH strategy is extremely powerful to test such channels.

In this letter, we reinterpret the results of the ALEPH search~\cite{ALEPH:2000vvi} to, for the first time, place a limit on ${\rm Br}(B_{s}\to{\rm invisibles})$.
As a side product, we also derive a bound on ${\rm Br}(B_{d}\to{\rm invisibles})$ that is only a factor of $\sim 6$ weaker than the current best limit derived with a dedicated search at BaBar~\cite{BaBar:2012yut}.
The power of the search strategy at ALEPH is manifested by the fact that the LEP data set contains less than one million $Z$ bosons decaying into $b\bar{b}$ versus the 471 million $B\bar{B}$ pairs contained in the BaBar sample.

It is also worth pointing out that other $B$ decay modes involving dark matter particles and other light neutral states can also be searched for using the ALEPH data sample. We refer to~\cite{Alonso-Alvarez:2021qfd} for an analysis of some of these modes and to~\cite{MartinCamalich:2020dfe,Ferber:2022rsf} for scenarios where these types of searches could cover yet uncharted parameter space. As a bonus, in Appendix~\ref{app:axion}, we provide the first bounds on $B\to \tau a$ and $B\to \rho a$ where $a$ is a light axion-like particle that does not interact with the detector.

\section{ALEPH search for B decays with large missing energy}

In what follows we give explicit details about our recast of the latest ALEPH search for $b$-quark decays with large missing energy at the $Z$ peak~\cite{ALEPH:2000vvi}. This search in turn was strongly based on~\cite{ALEPH:1992zwu,ALEPH:1994bih} -- see~\cite{Tomalin:806041} for many checks and details about the original analyses. 

\textit{Event Sample, Cuts and Efficiencies:} The event sample contains approximately 4 million hadronically decaying $Z$ bosons. Explicitly, according to Table 1.2 of~\cite{ALEPH:2005ab}, the number of hadronically $Z$ decays used in the legacy analysis is
\begin{align}\label{eq:NZhadrons}
    N(Z\to {\rm hadrons}) = 4.07\times 10^6\,.
\end{align}
For the analysis~\cite{ALEPH:2000vvi}, a tagging method is used to select signal-like events with rather high efficiency. 
Events are divided into two hemispheres defined by the plane perpendicular to the thrust axis. 
Candidate hemispheres with large missing energy are kept only if there is a positive $b$ tag in the other hemisphere (${\rm efficiency} = 0.88$). The $b$ tagged hemisphere is simply required to have $E_{\rm miss} < 25\,{\rm GeV}$ and to have at least six good tracks. 
Cuts are also performed on the value of the thrust to ensure that events are dijet-like and also on the direction of the thrust axis to ensure that they are well contained in the detector. 
Events containing a moderately energetic light lepton ($e/\mu$) in hemispheres with large missing energy are removed.
This is done to efficiently reject semileptonic $b\to \ell \bar{\nu}_\ell X $ decays, which have a large branching fraction and produce significant amounts of missing energy. 

After considering all the cuts and efficiency selections, the ALEPH collaboration reports the following signal efficiencies for their target signal modes:
\begin{align}
   {\rm efficiency}(B^-\to \tau^-\bar{\nu}_\tau) &= 8.1\,\%\,, \label{eq:efficiencY_btaunu} \\
   {\rm efficiency}(b\to s\bar{\nu}\nu) &= 8.8\,\%   \,. \label{eq:efficiencY_bsnunu}
\end{align}
Simulating the precise signal efficiency for the completely invisible $B_{d/s}$ decays that we are interested in is beyond the scope of this work. However, as discussed in the previous paragraphs, the efficiency is mostly dictated by selections on the hemisphere opposite to the signal one. This explains why the efficiencies in Eq.~\eqref{eq:efficiencY_btaunu} and Eq.~\eqref{eq:efficiencY_bsnunu} are very similar which should also be the case for our decays of interest. In consequence, in what follows we take\footnote{ALEPH data and its analysis tools are public, and there are ongoing efforts to translate it to modern formats which could allow in the future for a definite test of this assumption. See~\cite{ALEPH_open}, and~\cite{Badea:2019vey,Chen:2021iyj,Chen:2023nsi} for recent reanalyses of the data.}:
\begin{align}\label{eq:efficiency}
   {\rm efficiency}(B_{d/s}\to {\rm invisibles}) &= 8\,\%\,,
\end{align}
for the decay modes of interest in this work. 

\textit{Fragmentation Function:} To understand the observed missing energy distribution pattern, it is key to have a handle on the fraction of the beam energy that $b$-hadrons carry at the time of their decay. This fraction is always smaller than one as a result of the hadronization process. On average, at LEP, $b$-hadrons decay carrying a fraction of $\sim 70\%$ of the beam energy~\cite{DELPHI:2011aa}. The ALEPH collaboration performed a dedicated study of the fragmentation function~\cite{ALEPH:2001pfo} and found that the energy spectra of $B$ mesons was well described by the fragmentation function of Kartvelishvili et al.~\cite{Kartvelishvili:1977pi}. In what follows, we use the values of $E_{\mathrm{B}}/E_{\rm beam}$ resulting from the fit to the data displayed in Figure 5 of~\cite{ALEPH:2001pfo} using the Kartvelishvili et al.~function. 

\textit{Data Processing, Signal Region and Backgrounds:} In the ALEPH search, each $b\bar{b}$ dijet event is divided into two hemispheres defined by the plane perpendicular to the thrust axis. After the various cuts and event selections were made, events were listed in histograms of $E_{\rm miss}$ defined as approximately $E_{\rm miss} \simeq E_{\rm beam} - E_{\rm visible}$, where $E_{\rm beam} = \sqrt{s}/2$ and $E_{\rm visible}$ is the measured energy in the given hemisphere. The correction factor to this formula is small, see page 282 of~\cite{ALEPH:1993oeg}, and the missing energy spectrum resolution is expected to be $\sim 2.8\,{\rm GeV}$. Although the ALEPH data set was taken at slightly different energies around and at the $Z$ peak (see~\cite{ALEPH:2001pfo}), we take $\sqrt{s} = 91.2\,{\rm GeV}$ for all the events. This is justified not only because at least $\sim 50\%$ of the events were recorded at that energy, but also because data taking was roughly symmetrical around the $Z$ peak.
We thus do not expect this simplification to have any significant impact on our results.

The signal region used by the ALEPH collaboration consists of events with $E_{\rm miss} > 35 \,{\rm GeV}$. The main source of background in this region are $b\to \tau \nu X $ and $c/b\to \ell \nu X$ semileptonic decays in which the neutrino carries a large missing energy and the energetic lepton is not identified in the signal hemisphere due to limited selection efficiencies. Although events where the visible energy is miss-reconstructed could in principle also contribute to the background rate, Ref.~\cite{Tomalin:806041} shows that these types of events do not populate the signal region with $E_{\rm miss} > 35 \,{\rm GeV}$.

\begin{figure}[t]
    \centering
    \includegraphics[width=0.47\textwidth]{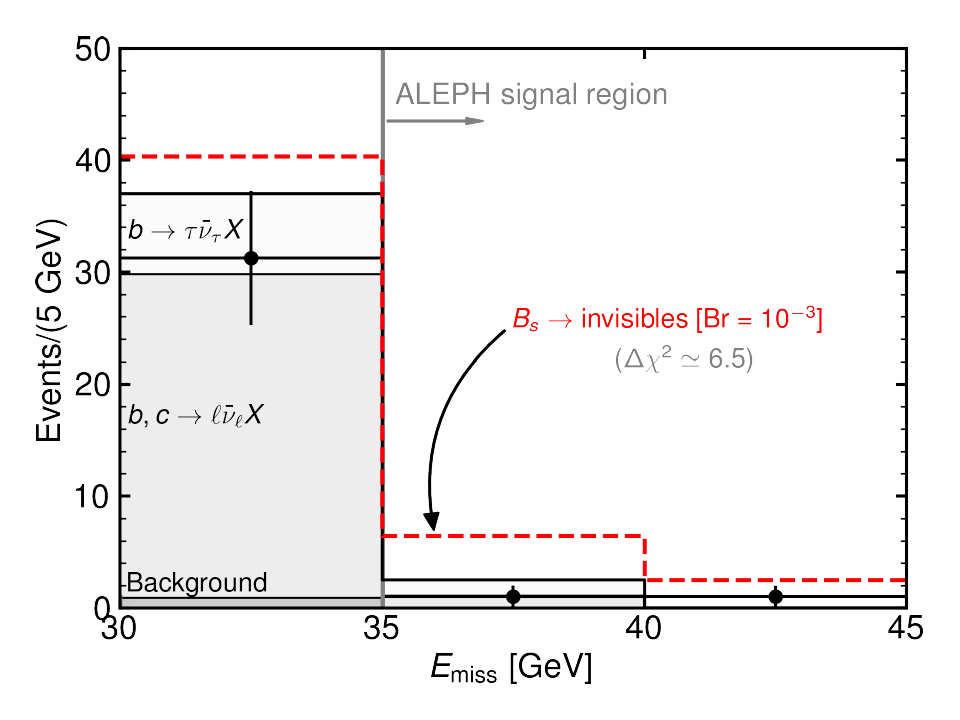} 
    \vspace{-0.3cm}
    \caption{Missing energy spectrum from $b$-flavored hadron decays at ALEPH~\cite{ALEPH:2000vvi}. The various bins contain background events where the missing energy is miss-reconstructed (dark grey), events from semileptonic $b$ and $c$ decays (grey), and $b\to \tau^- \bar{\nu}_\tau X $ (light grey). We superimpose (dashed red) the contribution from $B_s \to {\rm invisibles}$ with a branching fraction of $10^{-3}$. The visibly large effect in the signal region translates into it being statistically disfavoured ($\Delta \chi^2 \simeq 6.5$).}
    \label{fig:missingenergyspectrum}
\end{figure}

The observed and expected background events in the relevant bins are given in Table~5 of~\cite{ALEPH:2000vvi} and displayed in Figure~\ref{fig:missingenergyspectrum}. 
Following the original analysis, for our recast we use a single inclusive bin with
\begin{align}
    N^{\rm observed}_{E>35\,\mathrm{GeV}} &= 2\,, \label{eq:ALEPHdataobs}\\
    N^{\rm expected}_{E>35\,\mathrm{GeV}} &= 2.5\pm 1.6\,. \label{eq:ALEPHdataexp}
\end{align}

\textit{Fragmentation Ratios:} In order to derive bounds on $B_d$ and $B_s$ decays, we need to account for the probability of a $b$-quark hadronizing and leading to a weakly-decaying $B_d$ or $B_s$ in $Z$ decays. This has been extensively studied using a combination of measurements of the LEP experiments.
The latest averages from HFLAV are~\cite{HFLAV:2019otj,HeavyFlavorAveragingGroup:2022wzx}:
\vspace{-0.3cm}
\begin{subequations}\label{eq:fragmentationfractions}
\begin{align}
    f_{B_s} &= 0.101 \pm 0.008 \,,\\
    f_{B_d/B^\pm} &= 0.407 \pm 0.007  \,,\\
    f_{\rm b-baryons} &= 0.085 \pm 0.011 \,.
\end{align}
\end{subequations}
In what follows we will use the central values, as we have checked that propagating these uncertainties does not appreciably change our results. 

\textit{Simulation and Analysis:} We simulate the missing energy spectrum of $B \to {\rm invisibles}$ decays by folding in the $b$-quark fragmentation function taken from~\cite{ALEPH:2001pfo} and associating as missing energy all the energy carried by the given $B$ meson. In addition, since the calorimeters have a finite energy resolution (see above), we add to the simulated spectrum a random $\Delta E_{\rm miss} = 2.8\,{\rm GeV}$ Gaussian uncertainty. Our simulation shows that approximately $\sim \!\! 45\%$ of the invisible $B$ decay events fall in the signal region. 
The number of expected signal events can then be calculated based on the number of hadronic $Z$ decays in Eq.~\eqref{eq:NZhadrons}, taking into account that ${\rm Br}(Z \to \bar{b}b) = (15.12\pm 0.05)\,\%$ and ${\rm Br}(Z \to {\rm hadrons}) = (69.91\pm 0.06)\,\%$ ~\cite{ParticleDataGroup:2022pth}, and assuming an efficiency of 8\% for our decays of interest, following Eq.~\eqref{eq:efficiency}. 

In order to compare the simulated spectrum with the data, we build a Poisson likelihood for the number of observed events given an unknown signal and background rate. In addition, we add to the likelihood a Gaussian distribution for the background rate with fixed variance informed by the estimate in Eq.~\eqref{eq:ALEPHdataexp}.
Since no excess of events over the expected background is observed, we use the confidence levels method~\cite{Junk:1999kv,Read:2002hq} to place an upper limit on the signal rate\footnote{We have checked that using a profiled likelihood method leads to similar results.}.
All together, this translates into
\begin{align}
    {\rm Br}(B_s \to {\rm invisibles}) &< 5.6\times 10^{-4}\,\,\, [90\%\, {\rm CL}]\,,\label{eq:limitsBs} \\
    {\rm Br}(B_d \to {\rm invisibles}) &< 1.4 \times 10^{-4}\,\,\, [90\%\, {\rm CL}]\,,\label{eq:limitsBd}
\end{align}
which amounts to requiring less than $3.7$ expected signal events in the signal region (4.6 at $95\%$ CL). 
 
Given the improvements in the predictions of semileptonic heavy quark decays over the last decades one may be worried that the background modeling used by ALEPH is by now obsolete. While this is unlikely to be true because the events with large missing energy are not necessarily those at the kinematical end point (see~\cite{Grossman:1995gt}), one can derive a very conservative bound assuming $N_{E > 35\,{\rm GeV}}^{\rm expected} = 0$. Namely, that ALEPH was seeing only signal. By following the same procedure as before we find that the 90(95)\% CL limit on the number of signal events would be $5.3(6.3)$, with the upper limits in Eq.~\eqref{eq:limitsBs} and~\eqref{eq:limitsBd} being relaxed accordingly.

On the other hand, one can wonder whether our $B\rightarrow\mathrm{invisibles}$ search is impacted by the SM processes $B^\pm \to \tau \nu$ and $b\to s\nu\bar{\nu}$ that were the original targets of the ALEPH search.
We find that the constraints shown in Eq.~\eqref{eq:limitsBs} and Eq.~\eqref{eq:limitsBd} are slightly strengthened by considering these SM channels as background in the signal region.
Using the current measurement of the branching ratio of $B^- \to \tau^- \bar{\nu}_\tau$~\cite{ParticleDataGroup:2022pth} and the SM prediction for $b\to s\nu\bar{\nu}$ from~\cite{Altmannshofer:2009ma}, together with the ALEPH signal rate estimations~\cite{ALEPH:2000vvi}, we find these processes to contribute $+0.93$ events in our signal region. 
By adding those to the expected background, the 90(95)\% CL upper limit on the number of $B_{d/s}\to {\rm invisibles}$ events becomes 3.4(4.3), only $\sim$10\% stronger than our benchmark bound in Eqs.~\eqref{eq:limitsBs} and~\eqref{eq:limitsBd}.

\section{Discussion, Conclusions \& Outlook}

Motivated by the recent evidence for $B^+\to K^+ \bar{\nu}\nu$ transitions at Belle II~\cite{BelleII_BKnunu,BelleII_BKnunu2}, in this study we have derived new limits on fully invisible neutral $B$ meson decays.
For that, we have reinterpreted an old search carried out by the ALEPH experiment that looked into $b$-quark decay events with large missing energy at LEP~\cite{ALEPH:2000vvi}.
Reproducing the original analysis as faithfully as possible, our recast implies that ${\rm Br}(B_s \to {\rm invisibles}) < 5.6\times 10^{-4}$ and $ {\rm Br}(B_d \to {\rm invisibles}) < 1.4 \times 10^{-4}$, both at 90\% CL.

This limit on $B_d \to {\rm invisibles}$ is less than a factor of 6 times looser than the one reported by the BaBar collaboration from a dedicated analysis of 471 million $B\bar{B}$ pairs~\cite{BaBar:2012yut}.
Taking into account that the full ALEPH data sample only contains approximately $0.9$ million $b\bar{b}$ pairs, it is clear that the ALEPH search strategy is extremely powerful to look for $B$ meson decays featuring large missing energy. The reason behind it is its inclusiveness: aside from standard selection cuts, the search selects events based solely on large missing energy requirement, which results in an efficiency as high as $\sim 8\%$. This should be contrasted to the strategy at $B$-factories which requires at least partial tagging of the $B$ meson at the non-signal hemisphere yielding efficiencies of $\sim 0.17\%$~\cite{BaBar:2012yut}.

To the best of our knowledge (see~\cite{ParticleDataGroup:2022pth}), our derived limit on $B_s\to {\rm invisibles}$ is the first of its kind.
Characterizing this channel is particularly timely because the Belle II collaboration has recently reported 3.5$\sigma$ evidence for the $B^+\to K^+\bar{\nu}\nu$ decay~\cite{Belle-II:2023esi}.
This detection is intriguing as the implied rate is $2.7\sigma$ above the Standard Model expectation~\cite{Becirevic:2023aov}.
If confirmed, this will be the first indication of a $b \to s \bar{\nu}\nu$ transition. 
Clearly, such transition could also trigger the $B_s \to{\rm invisibles}$ process. 
The rate for this mode in the Standard Model is predicted to be minuscule~\cite{Lu:1996et,Badin:2010uh,Bortolato:2020bgy}, but extensions of the Standard Model featuring new light species could make it substantial. 
In fact, recent global analyses of $B \to K \bar{\nu} \nu$ modes show that one could expect branching ratios as large as ${\rm Br}(B_s \to {\rm invisibles}) \sim 10^{-5}-10^{-4}$~\cite{Bause:2023mfe,Felkl:2023ayn}. 
While the limit that we have derived in this work is not able to reach such sensitivities, it shows a potential new avenue to shed light on $b\rightarrow s$ transitions.

Looking forward, Belle II running at the $\Upsilon(5S)$ resonance with $5\,{\rm ab}^{-1}$ of luminosity is expected to reach sensitivities of ${\rm Br}(B_s\to {\rm invisibles})\lesssim 10^{-5}$~\cite{Belle-II:2018jsg}. Looking further ahead, our study demonstrates that an ideal place to look for this type of decays would be the Tera-$Z$ factory at FCCee~\cite{FCC:2018byv,FCC:2018evy}. 
With a total of $\sim 6\times 10^{12}$ $Z$ bosons, the sensitivity to invisible $B$ decays should be excellent.
Although a detailed forecast is beyond the scope of this paper, we expect branching ratios ${\rm Br}(B_s \to {\rm invisibles}) \sim 10^{-5}-10^{-4}$ to be fully covered in such an experiment. This will test new physics models capable of explaining the recent Belle II measurement~\cite{Felkl:2023ayn}. 
Even if the claimed detection were to be refuted, FCCee should have no problem reaching the Standard Model prediction for $b\to s\bar{\nu}\nu$ at the level of ${\rm Br} \sim 10^{-5}$.
Beyond there, disentangling a potential new physics signal from the SM contribution could be challenging, but one still expects rather fine sensitivities, as was seen in~\cite{Amhis:2021cfy,Fedele:2023gyi,Amhis:2023mpj} looking at similar types of $b$ decays at FCCee.

\begin{center}
{\bf\small ACKNOWLEDGEMENTS}
\end{center}
We would like to thank Ian Tomalin for very useful discussions on the original ALEPH searches, in particular, about the signal efficiency expected for invisible $B$ decays. We would like to thank also Enrico Nardi for useful discussions about the role of hadronization in inclusive $b\to s \bar{\nu}\nu$ decays. This work was supported by the Natural Sciences and Engineering Research Council (NSERC) of Canada.

\appendix
\section{Bounds on $B\to \rho \, a$ and $B^+\to \tau^+ \, a$}\label{app:axion}

A procedure similar to the one described above can be used to place limits on partially invisible flavor-violating $B$ meson decays involving light particles like an axion or a dark photon.
In particular, and due to the inclusive nature of the search strategy, limits on channels that have so far not been directly targeted by experiments, like $B\rightarrow \rho a$ (see Table I in~\cite{MartinCamalich:2020dfe}) and $B\rightarrow \tau a$, can be derived.
Though not directly related with the main aim of this work, in this appendix we report the limits for archival purposes.

\begin{figure}[h]
    \centering
    \includegraphics[width=0.47\textwidth]{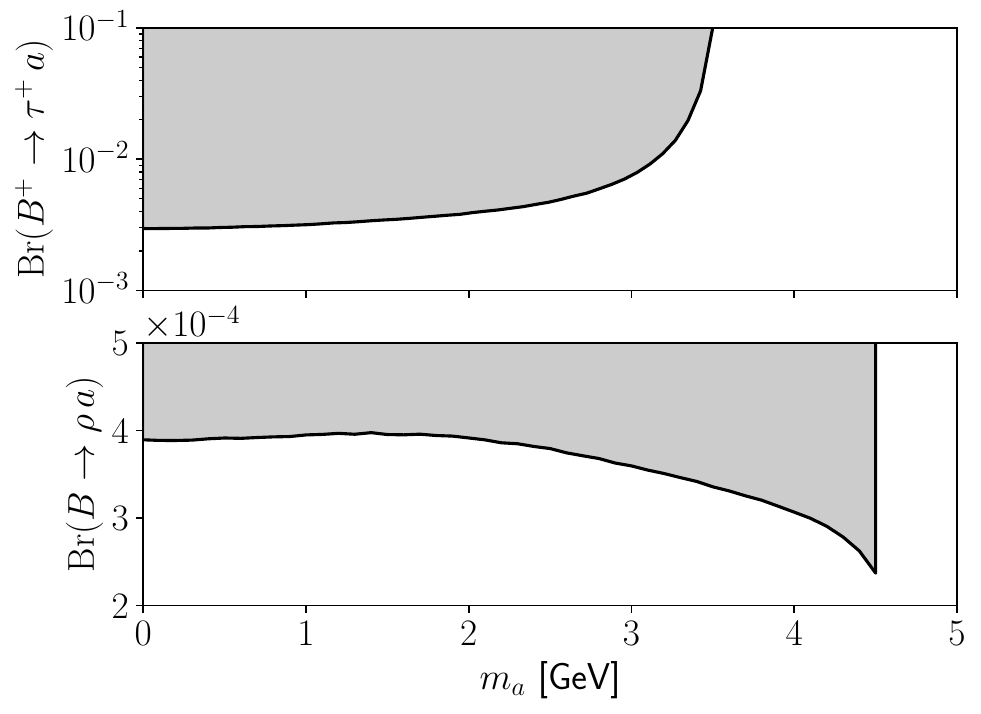} 
    \vspace{-0.3cm}
    \caption{Limits on flavor-violating $B$ meson two-body decays involving an axion from a recast of the ALEPH~\cite{ALEPH:2000vvi} analysis.}
    \label{fig:Br_B_a}
\end{figure}

Assuming that the $B$ meson decays isotropically and that the missing energy of the event is fully carried by the axion $a$, we obtain the fraction of events with $E_{\rm miss} > 35\,{\rm GeV}$ in the lab frame via a Monte Carlo simulation. The resulting branching ratio constraints as a function of the axion mass are shown in Fig.~\ref{fig:Br_B_a}.
For $B\rightarrow \rho \, a$, we include the charged $B^+\rightarrow \rho^+ \, a$ as well as neutral $B_d\rightarrow \rho_0 \, a$ channels.
For $B^+\rightarrow \tau^+ \, a$, we correct the limit by a factor of $1/{\rm BR}(\tau \to \mathrm{hadrons}\, +\, \nu)$ given that the ALEPH search rejects events with energetic electrons and muons such as those that arise from leptonic $\tau$ decays.
For a massless axion, we find
\begin{align}
    {\rm Br}(B \to \rho \, a) &<  3.9 \times 10^{-4}\,\,\, [90\%\, {\rm CL}]\,, \\
    {\rm Br}(B^+ \to \tau^+ \, a) &< 3.0 \times 10^{-3}\,\,\, [90\%\, {\rm CL}]\,.
\end{align}
Note that the limits on $B^+ \to \tau^+ \, a$ are quite conservative since for simplicity we do not count the missing energy carried away by the neutrinos in the $\tau$ decay chain. 

\newpage 
\bibliography{biblio}

\end{document}